\documentclass[aps,prl,preprint,superscriptaddress,nofootinbib]{revtex4-2}
\usepackage{graphicx}
\usepackage{dcolumn}
\usepackage{color}
\usepackage{bm}
\usepackage{amsmath}

\begin{document}
\title{Phase control of magnons in the van der Waals antiferromagnet NiPS$_3$\\}
\author{Shingo\ Toyoda*}
\affiliation{Department of Physics, University of California, Berkeley, California 94720, USA}
\affiliation{Materials Science Division, Lawrence Berkeley National Laboratory, Berkeley, California 94720, USA}
\author{Jonathon\ Kruppe*}
\affiliation{Department of Physics, University of California, Berkeley, California 94720, USA}
\affiliation{Materials Science Division, Lawrence Berkeley National Laboratory, Berkeley, California 94720, USA}
\author{Kohtaro\ Yamakawa}
\affiliation{Department of Physics, University of California, Berkeley, California 94720, USA}
\affiliation{Materials Science Division, Lawrence Berkeley National Laboratory, Berkeley, California 94720, USA}
\author{James\ Analytis}
\affiliation{Department of Physics, University of California, Berkeley, California 94720, USA}
\affiliation{Materials Science Division, Lawrence Berkeley National Laboratory, Berkeley, California 94720, USA}
\affiliation{CIFAR Quantum Materials, CIFAR, Toronto, Ontario M5G 1M1, Canada}
\author{Joseph\ Orenstein}
\affiliation{Department of Physics, University of California, Berkeley, California 94720, USA}
\affiliation{Materials Science Division, Lawrence Berkeley National Laboratory, Berkeley, California 94720, USA}
\date{received\hspace*{3cm}}

\hyphenpenalty=5000\relax
\exhyphenpenalty=5000\relax
\sloppy

\begin{abstract}

We demonstrate phase control of magnons in the van der Waals antiferromagnet NiPS$_3$ using optical excitation by polarized light. The sign of the coherent precession of spin amplitude changes upon (1) reversing the helicity of a circularly polarized pump beam, or (2) rotating the polarization of a linearly polarized pump by $\pi/2$. Because these two excitation pathways have comparable generation efficiency, the phase of spin precession can be continuously tuned from 0 to $2\pi$ by controlling the polarization state of the pump pulse. The ability to excite magnons with a desired phase has potential applications in the design of a spin-wave phased array and ultrafast spin information processing.
 \end{abstract}

\def\thefootnote{*}\footnotetext{These authors contributed equally to this work}

\maketitle

Coherent optical excitation of low energy collective modes is an important tool for both fundamental research and integration of materials into information technologies. One of the most prominent examples are magnons, the collective excitations of magnetic order \cite{kruglyak2010magnonics,barman20212021}. Magnons in insulators can propagate over macroscopic distances without Joule heating, making them promising for energy-efficient computing \cite{kajiwara2010transmission}. While recent years have seen rapid advances in all-electrical control of incoherent magnons \cite{kajiwara2010transmission,cornelissenmagnons,lebrunhematite1,lebrunhematite2}, these methods forego the potential of coherence. The phase information carried by coherent magnons provides an opportunity to design wave-based logic circuits that allow for non-Boolean computing, with the potential to operate faster and more efficiently than traditional Boolean logic circuits \cite{schneider2008realization,kostylev2005spin,chumak2015magnon,csaba2014spin}. In another application, tuning the relative phase of an array of spin wave emitters allows control over the direction of wavefront propagation. This enables the creation of a "spin-wave phased-array antenna" which is desirable for spin based information processing \cite{yoshimine2014phase,song2019omnidirectional}. 

While phase control of microwave-excited spin waves has been achieved through varying applied magnetic field \cite{schneider2008realization} or microwave amplitude \cite{ustinov2013nonlinear}, such techniques are limited to ferromagnetic resonances within the gigahertz frequency range. Aiming at higher operation speeds, the focus of magnonics research is shifting from ferromagnetic materials towards antiferromagnets, which have the potential for terahertz bandwidth applications \cite{jungwirth2016antiferromagnetic,baltz2018antiferromagnetic,nvemec2018antiferromagnetic}. Two-dimensional (2D) van der Waals (vdW) antiferromagnets have attracted attention as they may serve as atomically-thin magnonic devices \cite{huang2017layer,gong2017discovery,fei2018two,cai2019atomically,long2020persistence,khan2020recent,wang2020prospects}. Among them, the transition metal thiophosphates $M$PS$_3$ ($M$: Mn, Fe, Ni) are of particular interest because of their ultrafast spin dynamics \cite{afanasiev2021controlling,belvin2021exciton,mertens2022ultrafast}. While FePS$_3$ and MnPS$_3$ show easy-axis type antiferromagnetism \cite{lanccon2016magnetic,chandrasekharan1994magnetism,kurosawa1983neutron,ressouche2010magnetoelectric}, NiPS$_3$ is a $XY$-type easy-plane antiferromagnet \cite{joy1992magnetism,wildes2015magnetic,kim2019suppression}, offering promise as a platform to explore 2D magnetic phenomena such as the Berezinski-Kosterlitz-Thouless transition and long-distance spin transport via spin superfluidity \cite{seifert2022ultrafast,takei2014superfluid}. 

Optical pump-probe methods are powerful tools to excite and detect picosecond time-scale spin dynamics, and a number of antiferromagnetic resonances have been detected using these techniques \cite{satoh2010spin,bossini2016macrospin,satoh2015writing,baierl2016nonlinear,kalashnikova2007impulsive,kalashnikova2008impulsive,belvin2021exciton,afanasiev2021controlling,zhang2020gatetuneable}. Despite this intense study, reports of phase control of optically-excited spin waves are limited to ferromagnets, and in these reports only partial shifts in phase have been achieved \cite{yoshimine2014phase,chernov2018control}. Realizing full phase control from $0$ to $2\pi$ of light-induced antiferromagnetic magnons has remained a challenge.

In this Letter, we demonstrate full phase control of coherent magnon excitations in the vdW antiferromagnet NiPS$_3$, achieved by varying the helicity of the pump photons. The unique feature of NiPS$_3$ that enables optical control of spin wave phase is the coexistence of two excitation mechanisms with comparable efficiency: one arising from linearly polarized (LP) light and the other from circular polarized (CP) light. In both cases the generation mechanism is fully coherent; the sign of the spin wave amplitude flips upon reversing circular polarization from left to right and  rotating the linear polarization by $\pi/2$. This indicates that the excitation of magnons takes place through a non-thermal process, with the potential to control of spin-wave phase via the light-matter interaction.
 
\begin{figure}
\includegraphics*[width=8.6cm]{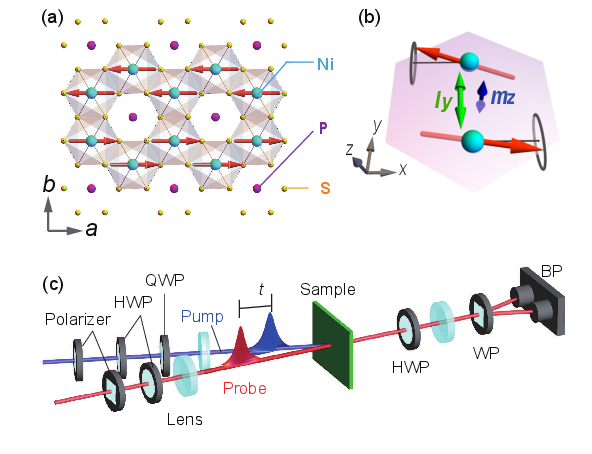}
\caption{
 (a) Crystal and magnetic structures of NiPS$_3$. Red arrows indicate the magnetic moments of Ni$^{2+}$ ions. (b) Schematic illustration of the oscillations of the Ni$^{2+}$ spins (red arrows), N\'{e}el vector (green arrow), and magnetization (blue arrow) of the in-plane magnon mode. (c) Schematic diagram of the experimental setup for the optical pump-probe measurement. HWP, QWP, WP, and BP denote the half-wave plate, quarter-wave plate, Wollaston prism, and balanced photo detector, respectively. 
\label{fig:001}
}
\end{figure}

NiPS$_3$ crystallizes in a monoclinic structure with the space group $C2/m$. The local magnetic moments are carried by the Ni$^{2+}$ ions with $S=1$ \cite{bernasconi1988lattice,brec1986review,ouvrard1985structural}. Trigonally distorted NiS$_6$ octahedra form an edge-sharing layered honeycomb lattice in the $ab$ plane, see Fig. 1(a). The trigonal axis is perpendicular to the $ab$ plane \cite{wildes2022magnetic}. Below the N\'eel temperature $T_{\rm{N}}$ = 155 K, Ni$^{2+}$ coplanar spins form ferromagnetic zig-zag chains along the $a$ axis that are antiferromagnetically coupled along the $b$ axis (Fig. 1(a)). The easy plane is slightly tilted from basal plane \cite{wildes2015magnetic,joy1992magnetism,wildes2015magnetic,kim2019suppression}. In the following, we use Cartesian coordinates $x\parallel \textbf{\textit{L}}$, $y\parallel b$, and $z\perp x,y$, where $\textbf{\textit{L}}$ denotes a N\'eel vector.

The spin wave dynamics of such a biaxial antiferromagnet, in which spins are strongly confined to a plane and weakly oriented within that plane, is characterized by two magnon modes \cite{rezende2019intro}. In the lower frequency mode the Néel vector $\bm{L}$ traces an ellipse with major axis in the magnetic easy plane (Fig. 1(b)) while the higher frequency mode the major axis is perpendicular to the easy plane. A previous pump-probe study of NiPS$_3$ showed that the lower frequency in-plane mode can be excited by a linearly polarized pump with a wavelength tuned to be resonant with a $d$-$d$ transition of Ni$^{2+}$ \cite{afanasiev2021controlling}.

We studied the dynamics of the in-plane mode of NiPS$_3$ as function of photon polarization using the setup shown schematically in Fig. 1(c). The tunable light source is an optical parametric amplifier (OPA) pumped by a regeneratively amplified laser, producing 100 fs pulses at 5 kHz repetition rate.  The OPA output, which was used as the pump, was tuned to 1240 nm to coincide with the $d-d$ transition from the ground state $^3A_{2g}$ to the excited state $^3T_{2g}$ of Ni$^{2+}$ ions. The probe wavelength was 800 nm. The polarization rotation of the transmitted probe light was detected using a Wollaston prism and balanced photodetector. 

Single crystals of NiPS$_3$ were grown by a chemical vapor transport method \cite{ho2021excitons}. Stoichiometric amounts of Ni, P, and S were mixed and the resulting powder mixture was loaded in a quartz tube. The ampoule was heated at 800$^{\circ}$ (heating zone) and 700$^{\circ}$ (growth zone) for a week. The obtained crystals were characterized by powder x-ray diffraction and Superconducting Quantum Interference Device (SQUID) magnetometry, and then mechanically exfoliated on a copper substrate with a hole of diameter 500 ${\mu}$m. The thickness of the sample was $\sim$10 ${\mu}$m. The sample was mounted in a He closed-cycled refrigerator.

\begin{figure}
\includegraphics*[width=12cm]{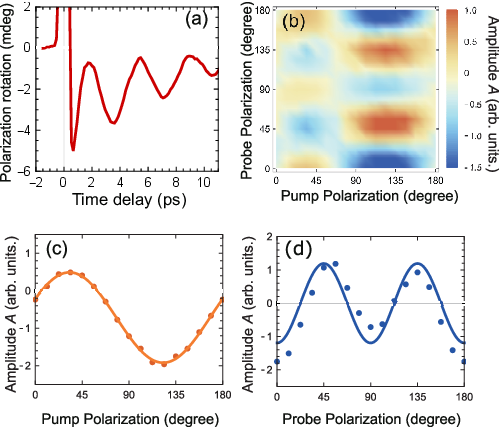}
\caption{
 (a) Time evolution of the probe polarization rotation for the linearly polarized pump measured at 12 K. (b) Probe and pump polarization dependence of the signed amplitude of linearly excited magnons in NiPS$_3$. Negative values of the amplitude correspond to opposite phase from positive values. Spin precession amplitude as a function of (c) pump polarization for the fixed probe polarization angle at 0 degree. (d) Probe polarization dependence of the magnon amplitude for the pump polarization at 135 degrees. The solid lines are best fits.
\label{fig:002}
}
\end{figure}

We first measured the properties of magnon generation and detection with linearly polarized light. Fig. 2(a) shows an example of photogenerated spin precession in NiPS$_3$ which can be fit to, 
\begin{equation}
    \Phi(t)=A \exp(-t/\tau)\cos (\omega t +\eta) + P_2(t),
\end{equation}
where $A$, $\omega$, $\tau$ and $\eta$ are an amplitude, frequency, decay time, and initial phase of the oscillations, respectively. $P_2(t)$ is a second-order polynomial that describes the smooth background on which the oscillations are superposed. An overview the dependence of the spin precession amplitude on pump and probe polarization is shown as a color map a function of polarization of the pump and probe beams is shown as a color map in Fig. 2(b). First, we note that the dependence on polarization demonstrates that the both the excitation and detection mechanisms is coherent, that is, magnons are generated and detected via a non-thermal light-matter interaction. Vertical and horizontal cuts through this plane clarify the dependence of $A$ on polarization direction. The horizontal cut plotted in Fig. 2(c) shows that $A\propto \sin2\theta$ at fixed probe angle, in this case $\theta=0^\circ$. Finally, the vertical cut at fixed pump polarization $\theta=135^\circ$ shows that in this case $A\propto \cos4\theta$. \cite{satoh2017excitation,afanasiev2021controlling}. Below we discuss how the dependence on the polarization of the pump and probe arise in an antiferromagnet with the symmetry of NiPS$_3$.

In the phenomenological model of the light-matter interaction the energy of the magneto-optical interaction is given by, \cite{satoh2017excitation,tzschaschel2017ultrafast,kalashnikova2008impulsive,gridnev2008phenomenological},
\begin{equation}
    {\cal W}_{MO}=-\frac{\chi_{ij}}{16\pi}{\cal E}_{i}{\cal E}^{*}_{j}{\cal T}(t),
\end{equation}
where ${\cal E}_{i}$ is the amplitude of the electric field of light. For light propagating in the $z$ direction we consider the $x$ and $y$ components, which have the form ${\cal E}_{x}\equiv {\cal E}_{0}\cos{\theta}$ and ${\cal E}_{y}\equiv {\cal E}_{0}\sin{\theta}e^{i \xi}$, respectively. Here $\theta$ is the azimuthal angle of rotation about the $z$ axis and $\xi$ is the relative phase of the $x$ and $y$ components of the electric field. LP corresponds to $\xi=0$, while CP corresponds to $\xi=\pm\pi/2$, $\theta=\pi/4$. The time-dependent function ${\cal T}(t)$ takes different forms depending on the type of excitation. For the displacive-type excitation \cite{hansteen2006nonthermal,atoneche2010large}, it is proportional to the Heaviside step function, while for the impulsive magnon excitation \cite{satoh2017excitation,kalashnikova2008impulsive}, it is proportional to Dirac delta function. The susceptibility tensor is given by
$\chi_{ij}=ik_{ijk}m_{k}+g_{ijkn}L_{k}l_{n}$ \cite{satoh2017excitation}, where the lowercase $l$ and $m$ are the light-induced changes in the N$\acute{\text{e}}$el vector and magnetization, respectively and the $L_{k}$ denote the components of the equilibrium Néel vector. When absorption is negligible, the Onsager reciprocity relation gives $\chi_{ij}(m,l,L)=\chi_{ji}^*(m,l,L)=\chi_{ji}(-m,-l,-L)$. Here we assume that this relation holds, because the pump and probe photon energies are below the parity-allowed charge transfer transition, leaving only the parity-forbidden weak $d-d$ transition. Consequently, the third rank axial tensor $k_{ijk}$ and fourth rank polar tensor $g_{ijkl}$ are purely real and satisfy $k_{ijk}=-k_{jik}$ and $g_{ijkl}=g_{jikl}$. The tensor components that are relevant to the in-plane mode are $k_{xyz}=-k_{yxz}$ and $g_{xyxy}=g_{yxxy}$, as the other components are either associated with the out-of-plane mode or vanish in the 2/m point group (see the Supplementary Information). The magneto-optical interaction energy of the lower frequency magnon mode can therefore be written as
\begin{equation}
\label{eq:lm_interaction}
    {\cal W}_{MO}
=-\frac{I}{16\pi}\sin2\theta\left[{\cal T}_m(t) k_{xyz} m_z \sin{\xi} + {\cal T}_l(t)g_{xyxy} L_x l_y  \cos{\xi}\right],
\end{equation}
where $I={\cal E}_{0}{\cal E}^{*}_{0}$ is proportional to the intensity of the pump light ($l_y$ and $m_z$ are sketched in Fig. 1(b). Eq. \eqref{eq:lm_interaction} shows that the $\sin(2\theta)$ dependence on pump polarization shown in Fig. 2 follows directly from the symmetry of NiPS$_3$. Note that in addition to coupling via linearly polarized light ($\xi=0$) to fluctuations in $l_y$, coupling via circularly polarized light ($\xi=\pi/2$) to $m_z$ is allowed. 

The detection of spin precession by the probe occurs via the inverse of the couplings described above and can arise from oscillations of $l_{y}$ and/or $m_{z}$. Fluctuations in $l_{y}$ give rise to magnetic linear birefringence, which has been shown \cite{satoh2017excitation} to manifest as the $\cos4\theta$ dependence on probe polarization that we observe (see Fig. 2(d)). Fluctuations in $m_{z}$ give rise to Faraday rotation which is independent of $\theta$. Hence, the detection of the in-plane magnon at the probe wavelength used in our measurements occurs via the magnetic linear birefringence associated with $l_{y}$ rather than the Faraday rotation mediated by $m_z$. 

Although the probe couples only $l_y$, the pump pulse can generate oscillations of both $l_y$ and $m_z$, via coupling to linearly and circularly polarization, respectively. The key to phase control of spin precession in NiPS$_3$ is that both the LP and CP modes of magnon generation are not only nonzero, but comparable in magnitude. This is illustrated in Figs. 3(a) and 3(b), which show time-resolved probe polarization rotation for LP, and CP pump pulses, respectively, plotted on the same time and amplitude scales. The comparable magnitude of precession generated by LP and CP pump photons is notable, as in other systems the relative efficiency of magnon excitation varies by orders of magnitude \cite{kalashnikova2008impulsive,tzschaschel2017ultrafast}. The spin precession frequency $\Omega$ is $0.28$ THz for LP and $0.27$ THz for CP, showing that the same spin wave mode is excited in both cases. Moreover, this frequency is consistent with the low-energy in-plane magnon mode found in the previous optical pump-probe \cite{afanasiev2021controlling} and electron spin resonance studies \cite{mehlawat2022low}. As illustrated in the Fig. 3(a) inset, the two responses to LP excitation shown are obtained by rotating the pump polarization by $\pi/2$; the resulting $\pi$ change in phase arises from the $\sin 2\theta$ term in the light-matter interaction ${\cal W}_{MO}$. Fig. 3(b) shows that the phase of the spin precession also changes by $\pi$ for CP pumping upon reversing the helicity, arising in this case from the $\sin \xi$ factor in ${\cal W}_{MO}$.  

In addition to the sign reversal on changing polarization, Fig. 3 also displays the initial phase, $\eta$ for each of the curves. As we discuss below, the initial phase reflects the nature of the magnon generation mechanism
\cite{gridnev2008phenomenological,kalashnikova2008impulsive,tzschaschel2017ultrafast}). The free energy of the easy-plane type antiferromagnet can be written as \cite{afanasiev2021controlling}
\begin{equation}
    F=\frac{1}{2M_0}(H_E{\bm M}^2+\frac{H_x}{2}(M_{x}^2+L_{x}^2)+\frac{H_z}{2}(M_{z}^2+L_{z}^2)) + {\cal W}_{MO},
\end{equation}
where $H_E$ and $H_{x,z}$ are the effective exchange field and anisotropy field, respectively. Dynamics described by Landau-Lifshitz equations,
\begin{equation}
    \frac{d{\bm m}}{d t}=-\gamma\{[{\bm M} \times{\bm H}^{eff}] + [{\bm L} \times{\bm h}^{eff}]\},
    \label{LLM}
\end{equation}
\begin{equation}
    \frac{d {\bm l}}{d t}=-\gamma\{[{\bm M} \times{\bm h}^{eff}] + [{\bm L} \times{\bm H}^{eff}]\},
    \label{LLL}
\end{equation}
where ${\bm H}^{eff}=- \partial F /\partial {\bm M}$ and ${\bm h}^{eff}= - \partial F/\partial {\bm L}$. 
Linearizing the equations leads to the following differential equations for the in-plane magnon mode, 
\begin{equation}
    \frac{d {m_z}}{\gamma d t}=- H_xl_y -h_l L_x {\cal T}_m(t),
    \label{eq:Mzdot}
\end{equation}
\begin{equation}
    \frac{d {l_y}}{\gamma d t}=H_0m_z + h_m L_x {\cal T}_l(t).
    \label{eq:Lydot}
\end{equation}
where $H_0\equiv H_x - H_z - 2H_E$ and $h_{m(l)}\equiv -\partial {\cal W}_{MO}/\partial m_z(\partial l_y)$.
Eqs. \eqref{eq:Mzdot} and \eqref{eq:Lydot} describe spin oscillation at frequency $\Omega=\gamma(H_0H_x)^{1/2}$ with normal coordinate $m_z=i(H_x/H_0)^{1/2}l_y$. The phase of the oscillation after the pulsed excitation is determined by the torque functions, ${\cal T}_{l,m}(t)$. To illustrate the dynamics after pulsed excitation at $t=0$, we consider the case of LP pump, where $h_m=0$. For the impulsive generation mechanism ${\cal T}(t) = \Delta t\delta(t)$ with $\Delta t$ being the laser pulse duration, integrating the equations of motion gives, $m_z(t)=-\gamma h_l L_x \Delta t \cos \Omega t$, whereas the displacive mechanism ${\cal T}(t) = \Theta(t)$ yields $m_z(t)=-\gamma h_l L_x\Omega^{-1} \sin \Omega t$. Hence, LP pump induces $\sin\Omega t$ ($\cos\Omega t$) time dependence of $l_y(t)$ for the impulsive (displacive) mechanism. In the case of excitation with CP light ($h_l=0$), Eqs. \eqref{eq:Mzdot} and \eqref{eq:Lydot} predict $l_y(t)\propto h_m \cos \Omega t$ and $l_y(t)\propto h_m \sin \Omega t$ for impulsive and displacive excitation, respectively.

Our experimental results show cosine-like time dependence for the LP pump (Fig. 3(a)) while the CP pump induces a superposition of sine and cosine dependence (Fig. 3(b)) of $l_{y}$. This indicates that the displacive-type excitation plays a dominant role for the LP pump, while both impulsive- and displacive-type excitation processes contribute to the generation of the in-plane magnon mode for CP pumping.  Coexistence of impulsive and displacive excitation mechanisms have been observed previously in both coherent phonon excitation \cite{garrett1996coherent} and coherent magnon excitation \cite{yoshimine2014phase}.

With this understanding of the magnon generation and detection mechanisms, we turn to the demonstration of phase control shown in Fig. 4. The polarization of the pump beam is tuned continuously between LP and CP by transmission through a half-wave plate (HWP) and quarter-wave plate (QWP), as shown in Fig. 1(c). Varying the angle of the HWP, with the QWP fixed, continuously tunes the helicity $\xi$ without changing the azimuthal angle $\theta$. Figure 4(a) shows time traces of the spin precession, for several $\xi$ at $\theta = 135^\circ$.  As the pump helicity is tuned, the phase of the oscillation shifts monotonically. Figures 4(b) and (c) display the phase and amplitude of the spin wave oscillation, obtained by fitting the traces to Eq.(1), as function of pump helicity. Fig. 4(b) demonstrates continuous tuning of the phase of the coherent magnon excitation from 0 to 2${\pi}$. 
\begin{figure*}
\includegraphics*[width=12.9cm]{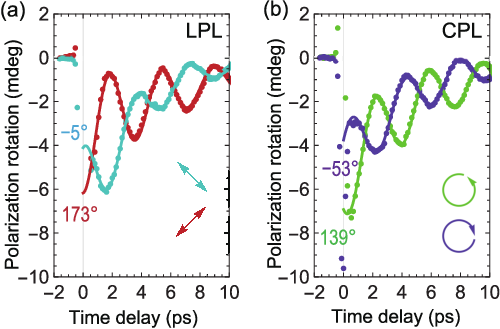}
\caption{
 (a) Change in the probe polarization rotation as a function of time delay via the excitation of (a) linearly polarized light and (b) circularly polarized light at 12 K. The arrows schematically illustrate the polarization state of the pump pulses. The solid lines are the fits to Eq. (1). The initial phases of the magnon oscillation obtained from the fittings are also shown. 
\label{fig:003}
 }
\end{figure*}
The dashed lines through the data points are obtained by summing the contributions from the CP and LP components of the elliptical pump pulses. The probe polarization rotation can be described by
\begin{equation}
\Phi(t)=(A_{LP}\cos{(\omega t + \eta_{LP})} \cos{\xi} + A_{CP}\cos{(\omega t + \eta_{CP})} \sin{\xi})\exp(-t/\tau),
\end{equation}
where $A_{LP}$ ($A_{CP}$) and $\eta_{LP}$ ($\eta_{CP}$) denote the amplitude and initial phase for the case of the LP (CP) pump, respectively. The parameters $\eta_{LP}= 180^{\circ}$ and $\eta_{CP}=140^{\circ}$ are obtained from measurements with pure LP and CP pump. Eq. 9 yields a resultant helicity dependent magnon oscillation, $A(\xi)\cos\left[\omega t+\eta(\xi)\right]e^{-t/\tau}$ where,
\begin{equation}
A(\xi)=\sqrt{A_{LP}^2\cos^2{\xi} + A_{CP}^2 \sin^2{\xi} - 2 A_{LP} A_{CP} \cos{\xi} \sin{\xi} \cos{\eta_{CP}}}
\end{equation}
and
\begin{equation}
\eta(\xi)=\tan^{-1}\frac{A_{CP} \sin{\xi} \sin{\eta_{CP}}}{-A_{LP} \cos{\xi} + A_{CP} \sin{\xi} \cos{\eta_{CP}}}.
\end{equation}
\begin{figure*}
\includegraphics*[width=12cm]{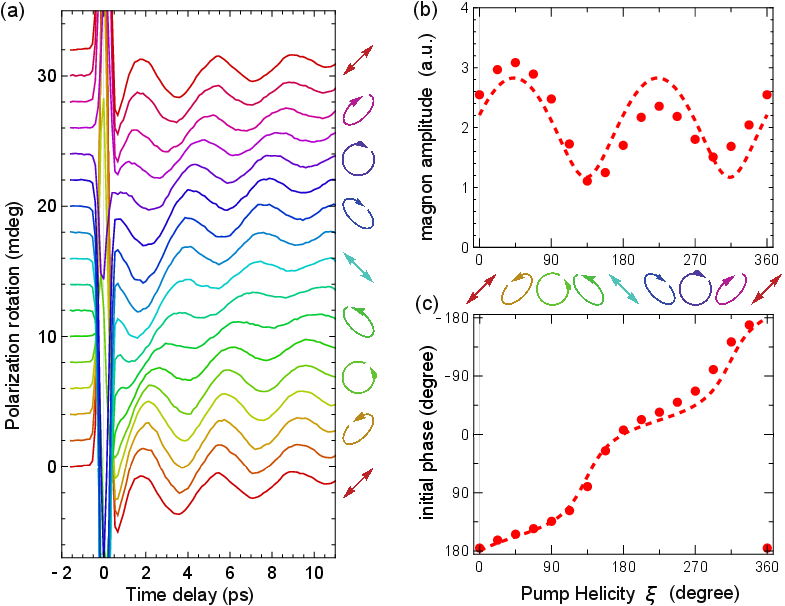}
\caption{
 (a) The time evolution of polarization rotation of the probe beam measured with varying polarization of the pump pulses at 12 K. The arrows schematically illustrate the polarization state of the pump. (b),(c) The pump polarization dependence of the (b) amplitude and (c) initial phase of the magnon oscillations. Dashed lines are fits to Eq. 10 and Eq. 11. 
\label{fig:004}
}
\end{figure*}

In principle, all magnetic systems in which magnons can be generated by both CP and LP light can manifest the phase control phenomenon described above. Indeed, both CP and LP excitation has been reported in other easy-plane type antiferromagnets, such as in FeBO$_3$ \cite{kalashnikova2008impulsive} and NiO \cite{tzschaschel2017ultrafast}. However, magnon phase control was not demonstrated, likely due to the fact that the magnon excitation by LP light is approximately two orders of magnitude more efficient than excitation by CP light \cite{kalashnikova2008impulsive,tzschaschel2017ultrafast}. 

To conclude, we performed time-resolved pump-probe measurements of the in-plane magnon mode in the antiferromagnet NiPS$_3$. The amplitude of spin precession was comparable for LP and CP pump processes, yet their initial phase is different. By tuning the pump helicity, the phase of the spin oscillation shifts from 0 to 2$\pi$ with little change in amplitude, thus establishing a reliable scheme for magnon phase control in the ultrafast regime. This technique paves the way for the construction of magnon-based devices such as spin-wave logic circuits \cite{schneider2008realization,kostylev2005spin} and phased array \cite{yoshimine2014phase,song2019omnidirectional}. Moreover, excitation of antiferromagnetic magnons can generate terahertz electromagnetic radiation \cite{belvin2021exciton}, and thus control the magnon phase may be useful for example as terahertz phased array antennas. 

We acknowledge support from the Quantum Materials program under the Director, Office of Science, Office of Basic Energy Sciences, Materials Sciences and Engineering Division, of
the U.S. Department of Energy, Contract No. DE-AC02-05CH11231. J.O received support from the Gordon and Betty Moore Foundation’s EPiQS Initiative through Grant
GBMF4537 at UC Berkeley. S.T. is supported by JSPS Overseas Research Fellowships.

\newpage
\bibliographystyle{apsrev4-2} 
\bibliography{bibliography} 

\newpage

\end{document}


\begin{center}
{\large {\textbf Supplementary material}}
\end{center}
\title{Full phase control of coherent magnon excitation in a van der Waals antiferromagnet NiPS$_3$\\}
\author{Shingo\ Toyoda*}
\affiliation{Department of Physics, University of California, Berkeley, California 94720, USA}
\affiliation{Materials Science Division, Lawrence Berkeley National Laboratory, Berkeley, California 94720, USA}
\author{Jonathon\ Kruppe*}
\affiliation{Department of Physics, University of California, Berkeley, California 94720, USA}
\affiliation{Materials Science Division, Lawrence Berkeley National Laboratory, Berkeley, California 94720, USA}
\author{Kohtaro\ Yamakawa}
\affiliation{Department of Physics, University of California, Berkeley, California 94720, USA}
\affiliation{Materials Science Division, Lawrence Berkeley National Laboratory, Berkeley, California 94720, USA}
\author{James\ Analytis}
\affiliation{Department of Physics, University of California, Berkeley, California 94720, USA}
\affiliation{Materials Science Division, Lawrence Berkeley National Laboratory, Berkeley, California 94720, USA}
\affiliation{CIFAR Quantum Materials, CIFAR, Toronto, Ontario M5G 1M1, Canada}
\author{Joseph\ Orenstein}
\affiliation{Department of Physics, University of California, Berkeley, California 94720, USA}
\affiliation{Materials Science Division, Lawrence Berkeley National Laboratory, Berkeley, California 94720, USA}

\date{received\hspace*{3cm}}

\hyphenpenalty=5000\relax
\exhyphenpenalty=5000\relax
\sloppy
\def\thefootnote{*}\footnotetext{These authors contributed equally to this work}

\setcounter{equation}{0}
\setcounter{figure}{0}
\setcounter{table}{0}
\setcounter{page}{1}
\makeatletter
\renewcommand{\theequation}{S\arabic{equation}}
\renewcommand{\thefigure}{S\arabic{figure}}
\renewcommand{\bibnumfmt}[1]{[S#1]}
\renewcommand{\citenumfont}[1]{S#1}
\maketitle
\section{Antiferromagnetic spin dynamics}
NiPS$_3$ is a XY-like easy-plane antiferromagnet with a small easy magnetic anisotropy within the plane. The spin-hamiltonian of the biaxial magnetic system is given by \cite{Wildes2022,01} 
\begin{equation}
    {\cal H}=\frac{1}{2}\sum_{<ij>} J_{ij} {\bm S_{i}} \cdot {\bm S_{j}} + D^{x} \sum_{i,j} (S_{i,j}^{x})^2  + D^{z} \sum_{i,j} (S_{i,j}^{z})^2,
\end{equation}
where $J_{ij}$ is exchange interaction between Ni$^{2+}$ ions, $D^{x}$ and $D^{z}$ represent the magnetic anisotropy strength along the $x$ and $z$ directions, respectively. With $J_{ij}>0$, $D^x<0$, $D^z>0$ and $|D^x|<|D^z|$, the Hamiltonian represents the easy-plane type antiferromagnet with the spin moments along the $x$-axis. We discuss the antiferromagnetic spindynamics using a macrospin approximation \cite{Rezende2019}. The magnetic moment of each sublattice is written as ${\bm M_{1,2}}=\gamma\hbar N{\bm S_{i,j}}$ with the number of spin per unit volume $N$ and the gyromagnetic constant $\gamma = g\mu_{B}/\hbar$. At the equilibrium, the sublattice magnetizations have the same magnitude $|{\bm M_{1}}|=|{\bm M_{2}}|=M_{0}$. The free energy of the magnetic-system per unit volume has a form
\begin{equation}
    F=\frac{H_E}{M_0}{\bm M_{1}} \cdot {\bm M_{2}} + \frac{H_{x}}{2M_0} (M_{1x}^2 +M_{2x}^2) +\frac{H_{z}}{2M_0} (M_{1z}^2 +M_{2z}^2).
\end{equation}
Here $H_E = SzJ/2\gamma \hbar$ and $H_{x,z}=2SD_{x,z}/\gamma \hbar$ are the effective exchange field and anisotropy field, respectively, where $z$ is the number of the nearest-neighbor sites. The magnetization vector ${\bm M}={\bm M_{1}}+{\bm M_{2}}$ and antiferromagnetic N\'{e}el vector ${\bm L}={\bm M_{1}}-{\bm M_{2}}$ are subject to constraints ${\bm M}\cdot{\bm L}=0$ and ${\bm M}^2 + {\bm L}^2=4M_0^2$. Hence, the free energy can be rewritten as
\begin{equation}
    F=\frac{1}{2M_0}(H_E{\bm M}^2+\frac{H_x}{2}(M_{x}^2+L_{x}^2)+\frac{H_z}{2}(M_{z}^2+L_{z}^2)).
\end{equation}
Here we neglect the constant term. The antiferromagnetic spin dynamics is described by Landau-Lifshitz equations
\begin{equation}
    \frac{\partial{\bm M}}{\partial t}=-\gamma\{[{\bm M} \times{\bm H}^{eff}] + [{\bm L} \times{\bm h}^{eff}]\},
    \label{LLM}
\end{equation}
\begin{equation}
    \frac{\partial {\bm L}}{ \partial t}=-\gamma\{[{\bm M} \times{\bm h}^{eff}] + [{\bm L} \times{\bm H}^{eff}]\},
    \label{LLL}
\end{equation}
where ${\bm H}^{eff}$ and ${\bm h}^{eff}$ are the effective magnetic fields, which are given by
\begin{equation}
    {\bm H}^{eff} = - \frac{\partial F}{\partial {\bm M}} = - \frac{H_{E} {\bm M}}{M_0} - \frac{1}{2M_0}\vectxyz{H_xM_x}{0}{H_zM_z},
\end{equation}
\begin{equation}
    {\bm h}^{eff} = - \frac{\partial F}{\partial {\bm L}} = - \frac{1}{2M_0}\vectxyz{H_xL_x}{0}{H_zL_z}.
\end{equation}
Linearlizing Eqs. \ref{LLM} and \ref{LLL}, we obtain following differential equations
\begin{equation}
    \dot{M}_z=-\gamma H_xL_y,
    \label{Mzdot}
\end{equation}
\begin{equation}
    \dot{L}_y= \gamma (H_x - H_z - 2H_e)M_z,
    \label{Lydot}
\end{equation}
and
\begin{equation}
    \dot{M}_y=\gamma (H_x - H_z)L_z,
\end{equation}
\begin{equation}
    \dot{L}_z= \gamma (-H_x + 2H_e)M_y.
\end{equation}
From these equations, we obtain the in-plane magnon mode and the out-of plane mode with the frequencies of $\omega_{IP}=\gamma \sqrt{H_x(H_x -H_z - 2H_e)}$ and $\omega_{OP}=\gamma \sqrt{(H_x-H_z) (H_x- 2H_e)}$, respectively. According to the inelastic neutron scattering study \cite{Wildes2022}, the dominant exchange interaction is between the third nearest-neighbor Ni$^{2+}$ ions $J = 13.5$ meV with $z=3$. The magnetic anisotropy are $D^x = -0.01$ and $D^z = 0.2$ meV in NiPS$_3$. Using these parameters, the frequency of the two magnon modes are estimated to be $\omega_{IP}=220$ and $\omega_{OP}=950$ GHz, which are in excellent agreement with the observed frequency in our pump-probe measurement.

\section{Light-matter interaction}
As discussed in the main text, the energy of the magneto-
optical interaction is given by,\cite{Kalashnikova2008, Gridnev2008, Pershan1966,02},
\begin{equation}
    {\cal W}_{MO}=-\frac{\chi_{ij}}{16\pi}{\cal E}_{i}{\cal E}^{*}_{j}{\cal T}(t),
\end{equation}
where 
\begin{equation}
    \chi_{ij}=ik_{ijk}m_{k}+g_{ijkn}L_{k}l_{n}.
\end{equation}
Here, we neglected the $m^2$ term because $|m|\ll|L|$. Since the $z$-component of the electric field of light is absent in our setup where the propagation direction of light is normal to the $xy$- ($ab$)-plane, we consider only the $x$ and $y$ components of the permittivity tensor, which are given by
\begin{equation}
    \chi_{xx}=g_{xxxx}L_{x}(L_{x}+l_{x})+g_{xxxz}L_{x}l_{z},
    \label{exx}
\end{equation}
\begin{equation}
    \chi_{yy}=g_{yyxx}L_{x}(L_{x}+l_{x})+g_{yyxz}L_{x}l_{z},
    \label{eyy}
\end{equation}
\begin{equation}
    \chi_{xy}=ik_{xyx}m_{x}+ik_{xyz}m_{z}+g_{xyxy}L_{x}l_{y},
    \label{exy}
\end{equation}
\begin{equation}
    \chi_{yx}=ik_{yxx}m_{x}+ik_{yxz}m_{z}+g_{yxxy}L_{x}l_{y}.
    \label{exy}
\end{equation}
The other components vanish under the point group of 2/m. For the in-plane mode oscillation ($l_{y}$ and $m_z$), only the $\chi_{xy}=ik_{xyz}m_{z}+g_{xyxy}L_{x}l_{y}$ and $\chi_{yx}=-ik_{xyz}m_{z}+g_{xyxy}L_{x}l_{y}$ have non-zero values.

\section{Magnon detection scheme}
 Hereafter we restrict our discussion to the in-plane magnon mode which involves $m_z$ and $l_y$ components. Since $m_z$ and $l_y$ contribute only to off diagonal components of the susceptibility (see Eqs. \eqref{exx} \eqref{eyy} \eqref{exy} ), the dielectric tensor can be written as
\begin{equation}
    \epsilon=
    \begin{pmatrix}
    \epsilon_{xx} & \delta\epsilon_{xy}\\
    \delta\epsilon_{yx} & \epsilon_{yy}
    \end{pmatrix}
    =
    \begin{pmatrix}
    \epsilon_{xx} & \delta\epsilon_{xy}^{s}+\delta\epsilon_{xy}^{a}\\
    \delta\epsilon_{xy}^{s}-\delta\epsilon_{xy}^{a} & \epsilon_{yy}
    \end{pmatrix},
\end{equation}
where $\delta\epsilon_{xy}^{s}=g_{xyxy}L_{x}l_{y}$ and $\delta\epsilon_{xy}^{a}=ik_{xyz}m_{z}$. The symmetric and antisymmetric components describe the magnetic linear birefringence and Faraday rotation, respectively. Our probe polarization measurement indicates that the polarization independent component (Faraday effect) is negligibly small. Therefore, we can reasonably assume that $\delta\epsilon_{xy}^{a}=0$ and the dielectric tensor is written as
\begin{equation}
    \begin{pmatrix}
    \epsilon_{xx} & \delta\epsilon_{xy}^{s}\\
    \delta\epsilon_{xy}^{s} & \epsilon_{yy}
    \end{pmatrix}.
\end{equation}
Diagonalizing the dielectric tensor, the refractive index and the eigen polarization are given by,
\begin{equation}
N_{1}=\sqrt{\epsilon_{xx}+\frac{(\delta\epsilon_{xy}^{s})^2}{\epsilon_{xx}-\epsilon_{yy}}}\ \   
    E_1=
    \left(
    \begin{array}{c}
    1\\
    \frac{\delta\epsilon_{xy}^{s}}{\epsilon_{xx}-\epsilon_{yy}}
    \end{array}
    \right),
\end{equation}
and
\begin{equation}
N_{2}=\sqrt{\epsilon_{yy}-\frac{(\delta\epsilon_{xy}^{s})^2}{\epsilon_{xx}-\epsilon_{yy}}}\ \ 
    E_2=
    \left(
    \begin{array}{c}
    \frac{-\delta\epsilon_{xy}^{s}}{\epsilon_{xx}-\epsilon_{yy}}\\
    1
    \end{array}
    \right).
\end{equation}
This corresponds to the rotation of the optical principle axis by an angle of $\beta=\frac{\delta\epsilon_{xy}^{s}}{\epsilon_{xx}-\epsilon_{yy}}$. The Jones matrix of the sample is given by,
\begin{eqnarray}
    S(\beta)=
    \begin{pmatrix}
    \cos{\beta} & -\sin{\beta}\\
    \sin{\beta} & \cos{\beta}
    \end{pmatrix}
    \begin{pmatrix}
    \exp{(i\phi_{xx}}) & 0\\
    0 & \exp{(i\phi_{yy}})
    \end{pmatrix}
    \begin{pmatrix}
    \cos{\beta} & \sin{\beta}\\
    -\sin{\beta} & \cos{\beta}
    \end{pmatrix} \nonumber \\  
    \approx
    \begin{pmatrix}
   \exp{(i\phi_{xx}}) & \beta [\exp{(i\phi_{xx}})-\exp{(i\phi_{yy}})]\\
    \beta [\exp{(i\phi_{xx}})-\exp{(i\phi_{yy}})] & \exp{(i\phi_{yy}})
    \end{pmatrix},\nonumber \\ 
    \   
\end{eqnarray}
where $d$ is the thickness of the sample, $\phi_{xx}=\frac{\omega d}{c}\sqrt{\epsilon_{xx}}$ and $\phi_{yy}=\frac{\omega d}{c}\sqrt{\epsilon_{yy}}$ describe the phase shift of the light waves in the sample for the light polarization along the $x$ and $y$ axis, respectively.
Let us discuss the probe polarization dependence of the detected signal of the spin precession by using a Jones matrix calculation. The Jones vector of the incident light to the sample is written as
\begin{equation}
    \left(
    \begin{array}{c}
    E_x\\
    E_y
    \end{array}
    \right)
    =
    \left(
    \begin{array}{c}
    \cos{\theta}\\
    \sin{\theta}
    \end{array}
    \right).
\end{equation}
The Jones matrix of the half wave plate between the sample and Wollaston prism can be written as
\begin{eqnarray}
    HWP(\theta)=
    \begin{pmatrix}
    \cos{(\frac{\theta}{2} + \frac{\pi}{8})} & -\sin{(\frac{\theta}{2} + \frac{\pi}{8})}\\
    \sin{(\frac{\theta}{2} + \frac{\pi}{8})} & \cos{(\frac{\theta}{2} + \frac{\pi}{8})}
    \end{pmatrix}
    e^{-\frac{i\pi}{2}}
    \begin{pmatrix}
    1 & 0\\
    0 & -1
    \end{pmatrix}
    \begin{pmatrix}
    \cos{(\frac{\theta}{2} + \frac{\pi}{8})} & \sin{(\frac{\theta}{2} + \frac{\pi}{8})}\\
    -\sin{(\frac{\theta}{2} + \frac{\pi}{8})} & \cos{(\frac{\theta}{2} + \frac{\pi}{8})}
    \end{pmatrix} \nonumber \\  
    =
    e^{-\frac{i\pi}{2}}
    \begin{pmatrix}
    \cos (\theta + \frac{\pi}{4}) & \sin (\theta + \frac{\pi}{4})\\
    \sin (\theta + \frac{\pi}{4}) & -\cos (\theta + \frac{\pi}{4})
    \end{pmatrix}.\nonumber \\ 
    \   
\end{eqnarray}
As a result the polarization vector before the Wollaston prism is written as,
\begin{eqnarray}
    \left(
    \begin{array}{c}
    E_x\\
    E_y
    \end{array}
    \right)&=&
    HWP(\theta)S(\beta)    \left(
    \begin{array}{c}
    \cos{\theta}\\
    \sin{\theta}
    \end{array}
    \right) \nonumber  \\ 
    &=&
    \left(
    \begin{array}{c}
    \cos{\theta}\cos{(\theta+\frac{\pi}{4})}e^{i \phi_{xx}}+\sin{\theta}\sin{(\theta+\frac{\pi}{4})}e^{i \phi_{yy}}+\beta[e^{i \phi_{xx}}-e^{i \phi_{yy}}]\sin{(\theta+\frac{\pi}{4}})\\
    \cos{\theta}\sin{(\theta+\frac{\pi}{4})}e^{i \phi_{xx}}-\sin{\theta}\cos{(\theta+\frac{\pi}{4})}e^{i \phi_{yy}}-\beta[e^{i \phi_{xx}}-e^{i \phi_{yy}}]\cos{(\theta+\frac{\pi}{4}})
    \end{array}
    \right).\nonumber \\
\end{eqnarray}
The signal of the balanced photo detector is the difference of the light intensity between two linearly polarized light
\begin{equation}
|E_x|^2-|E_y|^2=\frac{1}{2}[1-\cos{(\phi_{xx}-\phi_{yy})}]\sin{4\theta}+2\beta[1-\cos{(\phi_{xx}-\phi_{yy})}]\cos{4\theta}.
\end{equation}
Hence, the signal detected by the lock-in amplifier will be
\begin{equation}
I(\theta)\propto I_{\text{pump on}}-I_{\text{pump off}}=\frac{2g_{xyxy}L_{x}l_{y}}{\epsilon_{xx}-\epsilon_{yy}}[1-\cos{(\phi_{xx}-\phi_{yy})}]\cos{4\theta},
\end{equation}
which is proportional to $l_{y}$ and $\cos{4\theta}$. Hence, $l_{y}$ can be tracked by the balanced photo detector signal which shows the $\cos{4\theta}$ probe polarization dependence.

\section{Static Birefringence measurement}
\begin{figure}[b]
\includegraphics*[width=13cm]{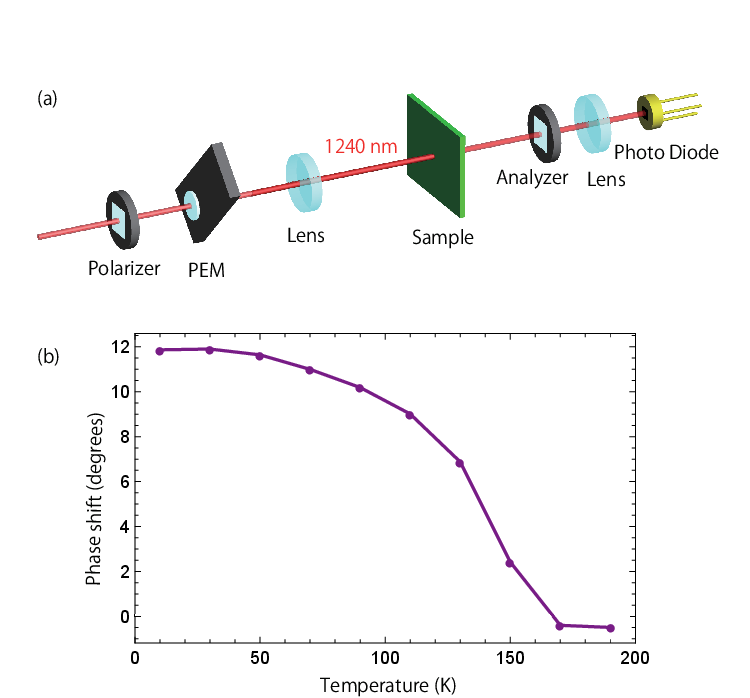}
\caption{
 (a) Experimental setup for the static birefringence measurements. PEM denotes a photoelastic modulator. (b) Temperature dependence of the phase shift between the ordinary and extraordinary polarization for the photon energy at 1.0 eV.
\label{fig:S001}
}
\end{figure}

In a birefringent medium, the polarization of pump pulses can vary depending on the position of the sample. The extent of the polarization change is influenced by the sample thickness. To accurately quantify the effect of birefringence on the pump polarization in our measurements, we employed a photoelastic modulator (PEM) in a transmission geometry (Fig. \ref{fig:S001}(a)) to measure the optical birefringence in NiPS$_3$ at the photon energy of 1.0 eV. The sample was illuminated by laser pulses from an optical parametric amplifier. The transmitted light intensity was measured by a photodiode (PD). The polarization of light was modulated with PEM (0.25 $\lambda$ mode), and the birefringence was measured by detecting the 1f signal using a lock-in amplifier. Figure \ref{fig:S001}(b) shows temperature dependence of the phase shift between extraordinary and ordinary light polarization. Based on the observed phase shift of approximately 12 degrees, one can reasonably assume that the pump beam retains its polarization as it passes through the sample. 
\clearpage
\newpage

